\documentclass[useAMS,usenatbib]{mn2e}
\usepackage{times}

\pdfoutput=1

\newcommand{\msun}{\rm \,M_{\odot}}
\usepackage{graphicx}
\usepackage{color}
\usepackage[T1]{fontenc}

\title[Infall of substructures onto a Milky Way-like dark halo]{Infall
of Substructures onto a Milky Way-like Dark Halo}

\author[Y.-S. Li \& A. Helmi]{Yang-Shyang Li\thanks{Email:
ysleigh@astro.rug.nl} and Amina Helmi\thanks{Email:
ahelmi@astro.rug.nl}\\
Kapteyn Astronomical Institute, University of Groningen, P.O. Box 800, 9700 AV Groningen, The Netherlands}
\begin{document}

\date{\today}   

\pagerange{\pageref{firstpage}--\pageref{lastpage}} \pubyear{2007}

\maketitle

\label{firstpage}

\begin{abstract}
We analyse the dynamical properties of substructures in a
high-resolution dark matter simulation of the formation of a Milky
Way-like halo in a $\Lambda$CDM cosmology. Our goal is to shed light
on the dynamical peculiarities of the Milky Way satellites. Our
simulations show that about 1/3 of the subhalos have been accreted in
groups. We quantify this clustering by measuring the alignment of the
angular momentum of subhalos in a group. We find that this signal is
visible even for objects accreted up to $z \sim 1$, i.e. 8 Gyr ago,
and long after the spatial coherence of the groups has been lost due
the host tidal field.  This group infall may well explain the ghostly
streams proposed by Lynden-Bell \& Lynden-Bell to orbit the
Milky Way.  Our analyses also show that if most satellites originate
in a few groups, the disk-like distribution of the Milky Way
satellites would be almost inevitable.  This non-random assignment of
satellites to subhalos implies an environmental dependence on whether
these low-mass objects are able to form stars, possibly related to the
nature of reionization in the early Universe.  With this picture, both
the ``ghostly streams'' and the ``disk-like configuration'' are
manifestations of the same phenomenon: the hierarchical growth of
structure down to the smallest scales.
\end{abstract}


\section{Introduction}
In cold dark-matter (CDM) cosmologies, large galaxies are the result
of the aggregation of smaller subunits. Some of these subunits may
survive until the present day in the form of satellites, while some
may be completely destroyed in the course of time, and contribute to
the field.  In $N$-body CDM simulations of the formation of
galaxy-size dark-matter halos, the satellites (often referred to as
substructures or subhalos) show significantly different properties
than those of the ``luminous'' satellites around galaxies like the
Milky Way.  An example of this discrepancy is the ``missing satellite
problem'': subhalos outnumber the bright satellites by factors of a
hundred or more \citep{kwg93,klypin99,moore99}. Furthermore, their
spatial distribution is typically much shallower than observed for the
luminous satellites \citep{gao04a,tbs04}. Therefore, the relation
between resolved substructures/subhalos in dark matter simulations and
the luminous satellites in galaxy halos is still unclear. Attempts to
reconcile these two populations, using semi-analytic models of galaxy
formation \citep{kwg93,benson02,kgk04} or full-fleshed SPH-simulations
\citep{maccio06,libeskind07} have produced interesting results, and
helped us gain insight into the relevant processes on the smallest
galactic scales. 
  
In the past ten years new attention has been drawn to the properties
of the satellite population in the Local Group. Starting with
\citet{lyndenbell95}, the existence of ghostly streams of satellites
(dwarf galaxies and globular clusters) was proposed. These objects
would share similar energies and angular momenta producing a strong
alignment along great circles on the sky \citep{palma02}.
Recently, it has been argued that the MW satellites define a disk-like
structure, that is so highly flattened (\textit{rms} thickness of
$10-30$ kpc) that may at first sight be inconsistent with CDM models
\citep{kroupa05,metz07}.  However, sophisticated modelling combining
semi-analytic galaxy formation recipes with dark matter simulations
has produced results that are consistent with observations
\citep{kang05,libeskind05,zentner05}.
 
In the past two years, about a dozen low surface brightness dwarf
satellite galaxies around the MW have been found in the Sloan Digital
Sky Survey
\citep{willman05a,willman05b,belokurov06,belokurov07,zucker06a,zucker06b,irwin07}. This
is an increase by a factor of two in the number of known satellites.
Because of the strong selection bias (SDSS focus is on the north
galactic pole), it is unclear whether these satellites confirm the
flattened disk-like structure. It is also unclear whether they help
solve the missing satellite problem \citep{sg07}. Satellites around M31 also show
peculiarities in their distribution, such as an excess of objects on
the side closest to the Milky Way \citep{mcc06}, and possibly a
similar degree of alignment \citep{kg06}.

These facts have motivated us to revisit the distribution and
properties of the subhalo population in CDM simulations. In
particular, here we focus on the infall of substructures onto a
Milky-Way like dark matter halo in a $\Lambda$CDM cosmogony. We use a
high-resolution dark-matter simulation that is a variant of the GA
series \citep{stoehr02,stoehr06} as described in
Section~\ref{simulation_sec}. We find evidence of group infall onto
the MW-like halo, which may explain the ghostly streams proposed by
\citet{lyndenbell95}. This is discussed in Section
\ref{group_infall_sec}. In Section~\ref{great_disk_sec} we show that
the disky configuration of satellites is consistent with CDM if most
satellites have their origin in a few groups. Thus both the ``ghostly
streams'' and the ``planar configuration'' are manifestations of the
same phenomenon: the hierarchical growth of structure down to the
smallest galactic scales. We present our conclusions in
Section~\ref{discussion_sec}.
%
%
%
\section{Subhalos in the dark matter simulations}

\subsection{Description of the simulations}
\label{simulation_sec}

We have analysed the GAnew series of high resolution simulations of a
MW-like halo \citep{stoehr06}.  The simulations were carried out with
GADGET-2 \citep{springel01}. The halo was selected from the M3
$\Lambda$CDM series \citep[see][]{csw03} with cosmological parameters
$\Omega_{0}=0.3, \Omega_{\Lambda}=0.7, h=0.7$, and Hubble constant
H$_{0}$=100$h$ km s$^{-1}$ Mpc$^{-1}$.  Candidate halos in this
simulation were selected according to the following criteria: {\it i)}
well-resolved, with $> 500$ particles; {\it ii)} environment similar
to the Local Group (free from nearby rich galaxy clusters); {\it iii)}
peak circular velocity approximately 220 km s$^{-1}$, and {\it iv)} no
major merger since $z =2$ \citep[see][]{gwn02}.  After choosing the
best MW-like halo, the zoomed initial conditions technique
\citep{tbw97} was used to produce a higher resolution simulation. The
MW-like halo was simulated four successive times with the mass
resolution increased by a factor $9.33$ each time.  In the highest
mass resolution simulation (GA3new), there are approximately $10^{7}$
particles within the virial radius.  Each of these re-simulations
produced 60 outputs equally spaced in $\log(1/(1+z))$ between $z=
37.6$ and $z=0$.  Table~\ref{summary_of_GAnew} summarises the
parameters and properties of the GAnew series.

\begin{figure*}
   \centerline{\includegraphics[width=1.03\textwidth]{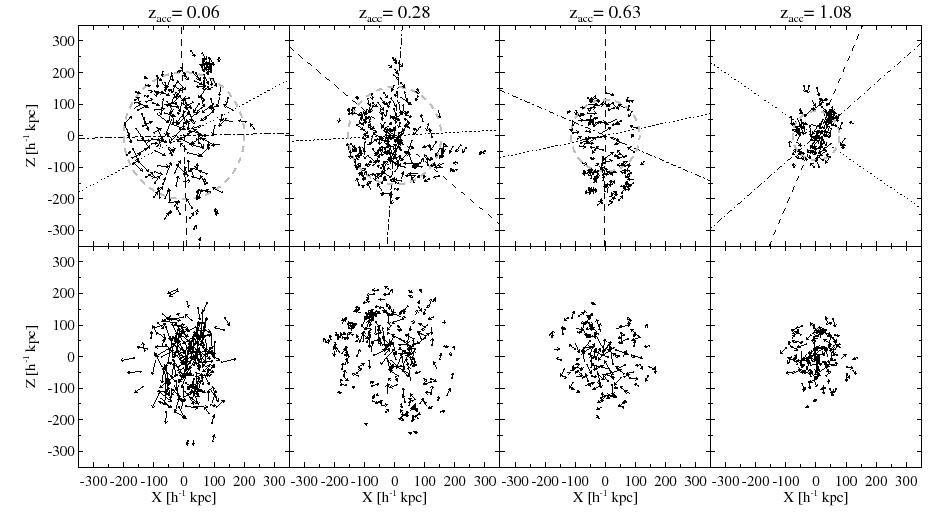}}
   \caption{Distribution of present-day subhalos in the Milky Way-like
halo in GA3new. The top panels show their distribution at the redshift
of accretion $z_{acc}$ in the principal axis reference frame defined
by the main halo at that epoch, while the bottom panels corresponds to
the present-day.  The arrows represent the velocity vector of each
subhalo: its length corresponds to $0.15$ of the velocity magnitude
while the orientation is defined by the direction of motion. The
dashed, dash dot and dotted lines in each panel indicate the
directions of the major, intermediate and minor axes of the ellipsoid
defined by these subhalos. The circle denotes the virial radius of the
host at the accretion epoch. Several groups of subhalos can be seen in
these diagrams, in particular at the time of accretion.}
\label{infall}
\end{figure*}

Virialised structures were identified in the high-resolution region of
the simulation using the standard friends-of-friends algorithm linking particles
separated by less than $0.2$ of the mean interparticle separation.
The {\footnotesize SUBFIND} algorithm \citep{springel01} was then
applied to each FOF group to find the gravitationally self-bound
subhalos. The smallest resolved subhalo contains 10 particles, and its
mass is $\sim 2.7 \times 10^{7}\msun$ in GA2new and $\sim 2.9
\times 10^{6}\msun$ in GA3new.  The number of subhalos present
in the final output is $504$ for GA2new and increases to $3,892$ for
GA3new. The virial mass of the MW-like halo is $2.4 \times
10^{12}h^{-1}\msun$ and the virial radius is $R_{vir}= 217\
h^{-1}$ kpc at present time in GA3new.

\begin{table}
\caption{Numerical parameters for the GAnew-series simulations. The
number of low-resolution particles remains at a roughly constant level
of $nlr \sim 1.272 \times 10^{6}$.} 

\label{summary_of_GAnew} 
  \begin{tabular}{ccrcl}
  \hline
  \hline
Name   &    $m_{\mathrm{p}}$ [$h^{-1}\msun$]  & N$_{\mathrm{HR}}$ & z$_{\mathrm{start}}$  & $\epsilon$ [$h^{-1}$ kpc]  \\
\hline
GA0new & 1.677$\times 10^{8}$ &    68\ 323 &    70 &  1.4 \\
GA1new & 1.796$\times 10^{7}$ &   637\ 966 &   80 &  0.8 \\
GA2new & 1.925$\times 10^{6}$ &  5\ 953\ 033 &   90 & 0.38 \\
GA3new & 2.063$\times 10^{5}$ & 55\ 564\ 205 &   60 & 0.18 \\
\hline
  \end{tabular}
\end{table}
%
%
%
\subsection{Group Infall of Dark Matter Substructures}
\label{group_infall_sec}

\subsubsection{Infall pattern}

The bottom panels of Fig.~\ref{infall} show the present-day spatial
distribution of subhalos accreted at four different redshifts
($z_{acc}$) in GA3new. We use the most bound particle of a subhalo to
represent its position and velocity. The accretion epoch of a subhalo
is determined as the time when its most-bound particle becomes part of
the FOF group in which the MW-like halo is located.  The reference
frame used in Fig.~\ref{infall} is defined by the principal axes of
the MW-like halo. Its principal axes have been determined by
diagonalising the inertia tensor
\begin{equation}
I_{ij}=\sum_{\mu} x_{i}^{\mu}x_{j}^{\mu}/\zeta^{2}_{\mu},
\label{inertia_tensor_ell_eq}
\end{equation}
where $x_{i}^{\mu}$ is the $i$ coordinate of the $\mu$th particle with
respect to a reference frame centred on the main halo,
$\zeta_{\mu}^{2} =
(y_{1}^{\mu})^{2}+(y_{2}^{\mu}/s)^{2}+(y_{3}^{\mu}/q)^{2}$, and
$y_{i}^{\mu}$ are its coordinates in the principal axes frame
\citep{dubinski,zentner05}.  We only consider particles with
$\zeta^{2} \le R_{vir}$, and iterate until $s$ and $q$ have changed by
less than $\sim 10^{-3}$.  The present time minor-to-major axis ratio
of the MW-like halo is $q=0.6$, and it has changed by less than 5\%
over the last 3 Gyr.

The top row of Fig.~\ref{infall} shows the spatial distribution of
subhalos at the time they were accreted onto the MW-like halo.  The
(grey) dashed circles in top panels show the virial radius at that
epoch.  There are $789, 231, 133$ and $117$ subhalos which have survived
until present time that were accreted at $z=0.06, 0.28, 0.63$ and
$1.08$ respectively.  Note that in the first column, we plot for
clarity only the first $200$ most massive subhalos accreted at
$z_{acc}=0.063$.

The distribution of subhalos is elongated roughly along the direction
of the major axis of the MW-like halo at each epoch.  The projected
major axis orientations of subhalos accreted at a given epoch are
indicated by the dashed lines in Fig.~\ref{infall}.  These are
computed using Eq.~(\ref{inertia_tensor_ell_eq}) but without
resampling\footnote{When the number of subhalos is small, as in the
right panel of Fig.~\ref{infall}, an iterative procedure quickly
leads to a sample with too few points, and hence to unreliable
results.}.  The alignment between the host and the subhalo's major
axes is prominent except for the $z_{acc}=0.28$ snapshot. At this time
the accreted subhalos are much more isotropically distributed,
preventing a clean determination of the orientation of the principal
axes of inertia (since the intermediate-to-major axis ratio is 0.98).

Fig.~\ref{infall} also shows that the distribution of subhalos is
clumpy on small scales. These clumps are formed by subhalos sharing
similar velocities (as shown by the arrows in this figure). This
indicates that infall does not occur in isolation but in groups. This
is very reminiscent of the way that clusters of galaxies grow through
the mergers of groups, residing in the intersections of filaments
\citep[e.g.][]{knebe04}. What we are seeing here is that also
galaxy-size halos grow via accretion of ``subgroup''-size structures.

The clumps of subhalos seen in Fig.~\ref{infall} share essentially
the same angular momentum at the time of infall.  This implies that,
even if the spatial clustering is less prominent after a few orbits
inside the main halo, the lumpiness may still be present in the space
of angular momenta, provided these are nearly conserved.

We quantify the degree of clustering by computing the two-point
``angular correlation function'', $\omega({\alpha})$, of the
present-day angular momentum of our subhalos. The angle $\alpha$ is
defined by the relative orientation of the angular momenta of any two
subhalos, i.e.  $\cos \alpha_{ij} = \mathbf{L}_{i} \cdot
\mathbf{L}_{j}/(|\mathbf{L}_{i}||\mathbf{L}_{j}|$). Therefore the
correlation function $\omega$ measures the number of pairs with
$\alpha_{ij} < \alpha$ compared to the expectations of an isotropic
distribution. To compute the expected number of random pairs we
average over $1,000$ realisations of a uniform distribution on the
sphere, whose size is given by the number of ``observed'' data points.
Note that any small scale clustering in angular momentum such as
observed in Fig.~\ref{infall} should manifest itself as an excess of
pairs with small angular separations.
 
\begin{figure}
\centerline{\includegraphics[width=0.5\textwidth]{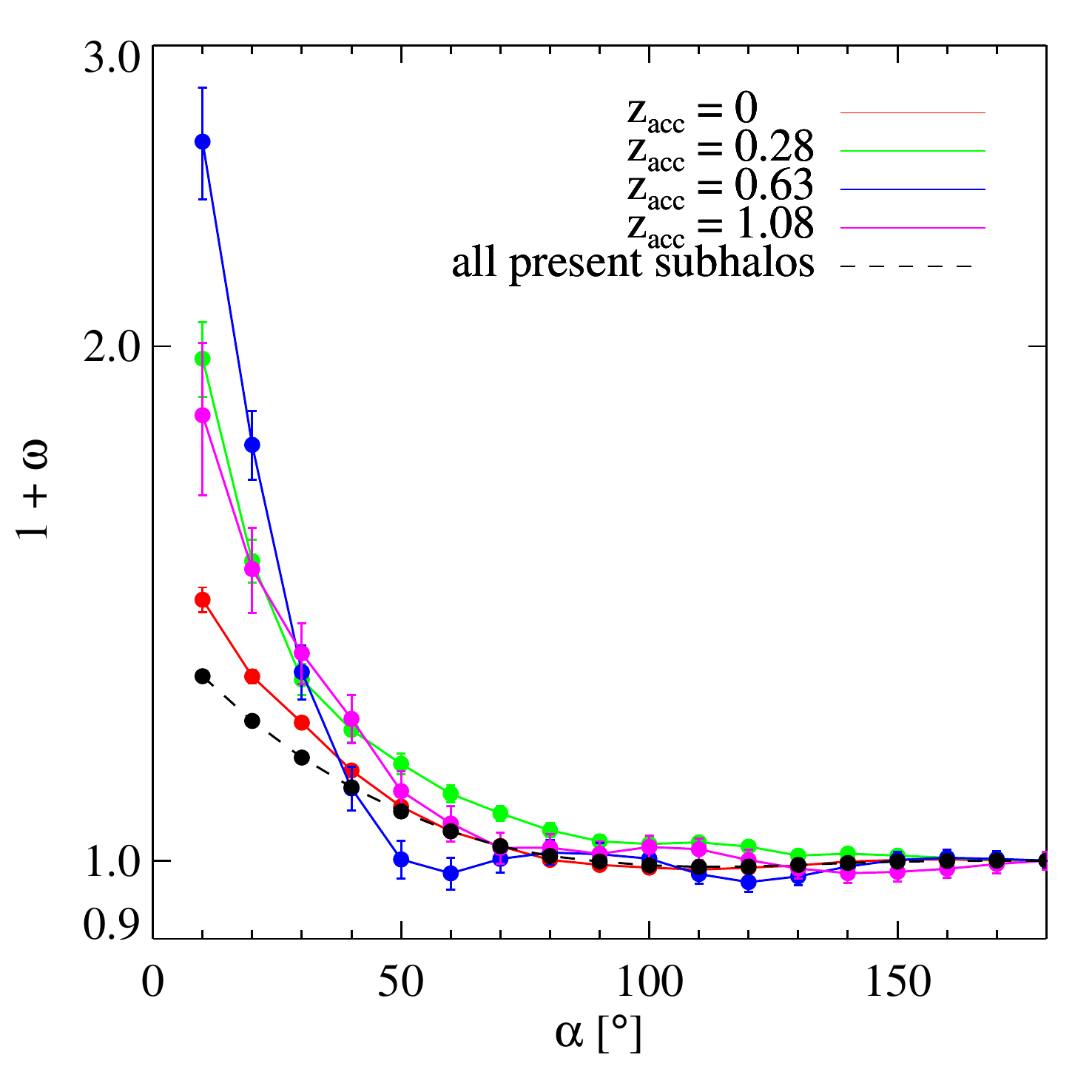}}
   \caption{Two-point ``angular correlation function'' using the
   present time angular momentum orientation of subhalos accreted in
   the last 8 Gyrs.  The excess in the first bins is indicative of the
   group infall and shows that this signal can persist for a very long
   time.}
\label{2point_corre_fn}
\end{figure}

Fig.~\ref{2point_corre_fn} shows the two-point ``angular correlation
function'' computed using the present-day angular momentum of subhalos
in our simulations. Different colours correspond to subhalos accreted
at different epochs.  We see a clear excess of pairs with angular
momentum orientation separation less than $10^{\circ}$ (and up to
$30^{\circ}$) compared to random samples.  This implies that the
signature of group infall is preserved in angular momentum even after
many Gyrs of evolution.  This signal is still discernible even for
subhalos accreted at $z \sim 1$.  The correlation function calculated
with all surviving subhalos at present is showed as the (black) dashed
line.
\subsubsection{Properties of groups}

We now focus on the characteristics of the groups accreted at the
various epochs. To identify groups we link pairs of infalling halos
whose angular momentum orientations are separated by $\alpha <
10^\circ$ and with relative distances $d < 40$ kpc at the time of
accretion. We found that this combination of $\alpha$ and $d$ values
results in a robust set of groups, maximizing their extent while
minimizing the number of spurious links.

\begin{figure}
      \centerline{\includegraphics[width=0.5\textwidth]{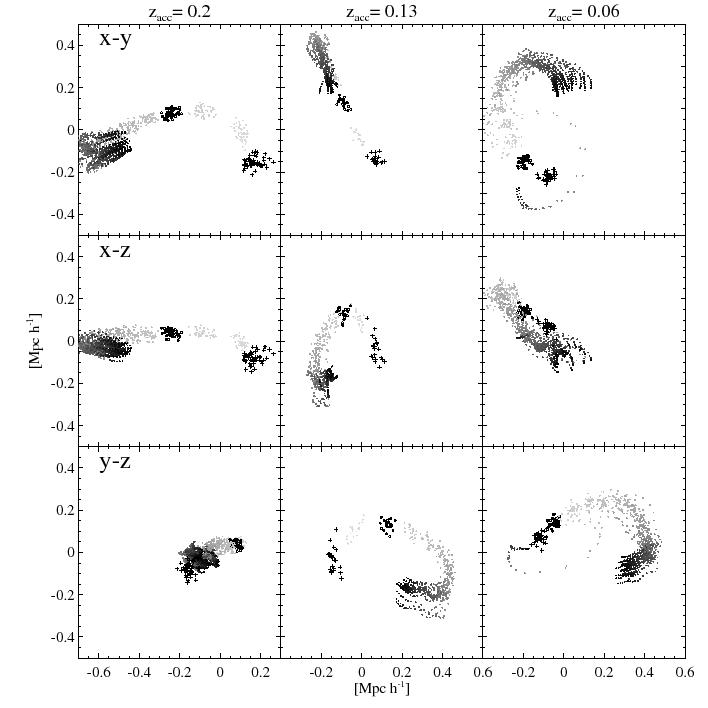}}
      \caption{Three examples of the trajectories of groups of
subhalos accreted at different epochs in the GA3new simulation
reference frame. These are some of the most abundant groups ever accreted.
The colour gradients indicate the arrow of time, from dark at high
redshift to light grey at the present. The positions at accretion
and present time are highlighted with open circles and crosses
respectively. There is some evidence that these groups themselves are
the result of the mergers of smaller groups, this is especially clear
for the group in the right-most panel.}
\label{orbital_plot}
\end{figure}

We follow the orbits of the groups identified from redshift $z \sim
4.2$ until present time.  Fig.~\ref{orbital_plot} shows the
trajectories of some of the richest groups of subhalos, which were
accreted 2.43, 1.65 and 0.84 Gyrs ago respectively.  Each dot
represents the position of a subhalo colour coded from high-redshift
(dark) to the present (light-grey).  The crosses correspond to the
present-day positions while those at the time of accretion are shown
as open circles.  Fig.~\ref{orbital_plot} clearly shows that the
groups of subhalos follow nearly coherent orbits as early as $z \sim
4.2$, long before the time of accretion.

\begin{figure}
   \centerline{\includegraphics[width=0.51\textwidth]{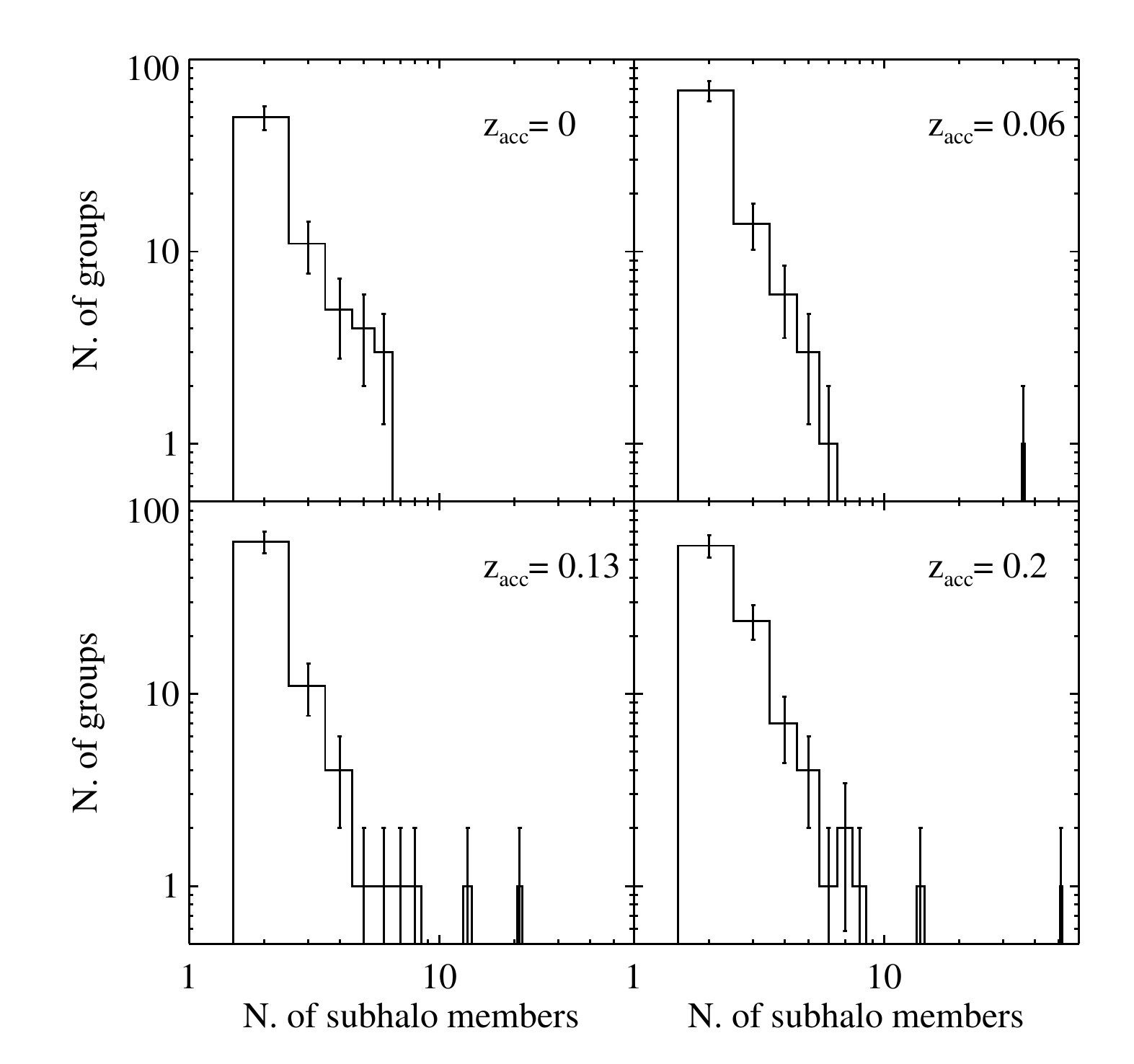}}
   \caption{Number counts of groups as function of the number of
member subhalos at four accretion epochs plotted in log-log scale. The
error bars are Poissonian.}
\label{grp_richness}
\end{figure}

\begin{figure}
   \centerline{\includegraphics[width=0.51\textwidth]{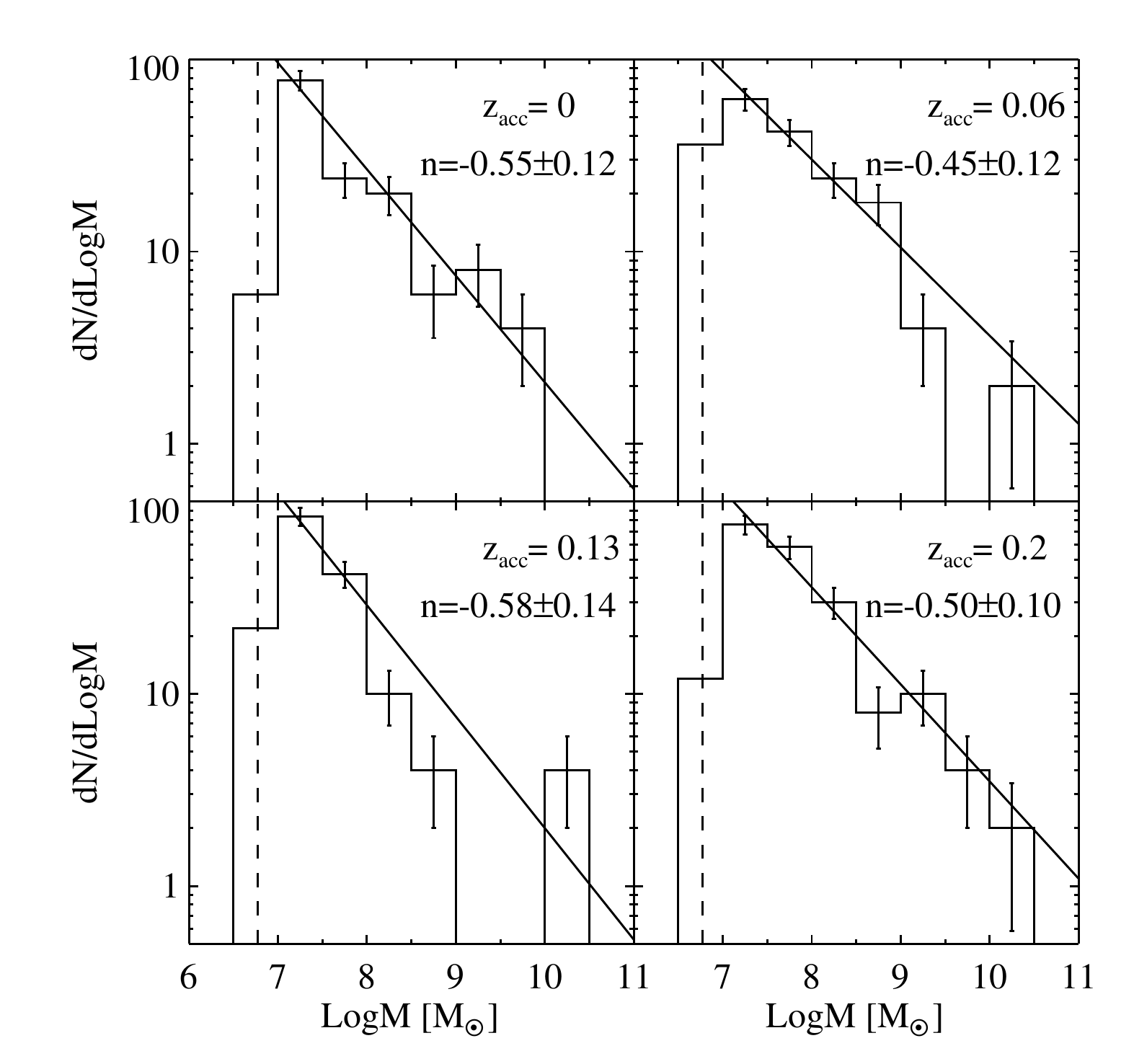}}
   \caption{Differential mass function of the groups shown in
   Fig.~\ref{grp_richness}. The power-law behaviour of the group
   mass function is similar to that of the subhalo mass function,
   albeit with a shallower slope.}
\label{massfn_allgrp}
\end{figure}

The characteristic size of the groups can be measured by computing the
number function of groups, i.e. how many groups have a given
number of subhalos.  Fig.~\ref{grp_richness} shows the number
function of groups accreted in the four most recent snapshots: present
time, 0.84, 1.65, and 2.43 Gyrs ago.  As can be seen from this Figure,
the shape is quite similar at all times, and most of the groups have a
small number of members.

Fig.~\ref{massfn_allgrp} shows the differential mass function of the
groups in Fig.~\ref{grp_richness} down to our resolution limit (the
dashed line, which corresponds to $\sim 5.89 \times 10^{6}\msun$).
Once again we find very similar power-law shapes for the mass
functions at different epochs.  This power-law shape is reminiscent of
the differential mass function of subhalos in cluster and galaxy-size
dark matter halos. The slope of the fitted $dN/d\log M \propto M^{n}$
relation is $n \sim -0.5 \pm 0.2$. Note that this is somewhat
shallower than that found for subhalos, where $n\sim -0.8 \pm 0.1$
\citep{stoehr03,delucia04,gao04b}.  This could well be due to insufficient mass
resolution: the fact that we are not resolving subhalos below $2.9
\times 10^6 \msun$, implies that many subhalos are accreted in
isolation, instead of in pairs or in groups. This effect is much
stronger at the low mass end of the group mass spectrum. For example,
a group with total mass $\sim 10^9 \msun$ can consist of ten
subhalos of $\sim 10^8 \msun$ or two of $\sim 5 \times 10^8
\msun$.  On the other hand a group of $\sim 10^7 \msun$ can only
be the result of a pair of subhalos of $5 \times 10^6 \msun$ in our
simulation. 

Our limited resolution also prevents us from quantifying the mass
function inside the groups. Nevertheless, and for our largest groups
we find that these are dominated by a few massive subhalos and many
small ones.

The group infall that we have been detecting in our simulation may
well be related to the ghostly streams reported by
\citet{lyndenbell95}.  The presence of satellites (dwarf galaxies and
globular clusters) sharing a common orbital plane seems rather
plausible in the context discussed here.  Instead of the disruption of
a large progenitor \citep{lyndenbell95} or the tidal formation of
satellites within gas-rich major mergers \citep{kroupa97}, we would be
witnessing the disruption by the tidal field of the Milky Way of a
``sub-group''-size object composed by dwarf galaxies.  The possible
implications of this finding are discussed in the Conclusions.
\subsubsection{Link to the Environment}

\begin{figure}
   \centerline{\includegraphics[width=.5\textwidth]{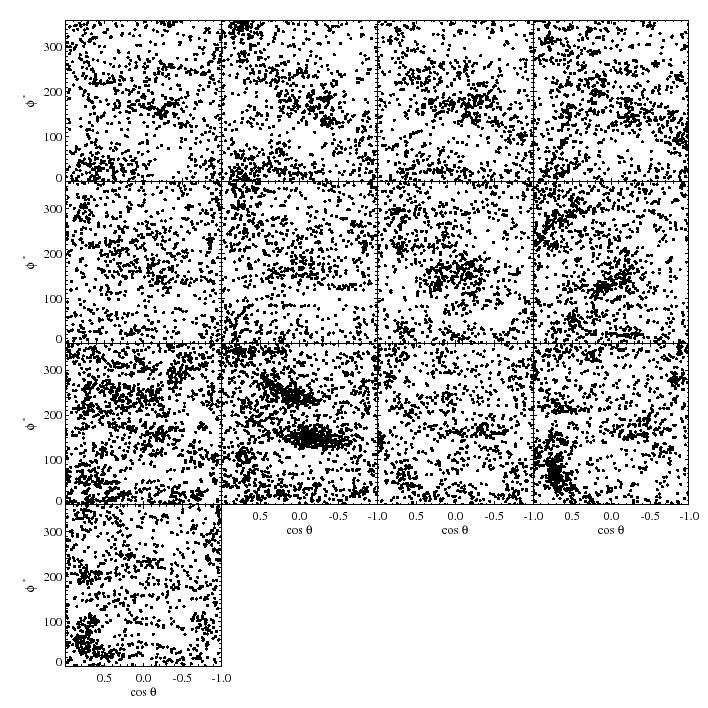}}
   \caption{Distribution of angular momentum orientation for the
subhalos accreted in the last 13 snapshots, i.e since $z = 1.08$.  The
arrow of time in this figure goes from right to left and from bottom
to top, i.e. the bottom left panel corresponds to $\sim 8$ ago, while
the top left to the present-day. Note the small scale structure
indicative of group infall as well as the large-scale pattern
associated to the filamentary structure of the tidal field. This
figure shows that not just one filament is actively feeding material
at any given time.}
\label{dist_of_angularmoment_orientation_accminus1}
\end{figure}

The present-time distribution of angular momentum orientations of
subhalos reflects both the anisotropy of the accretion pattern and the
dynamical processes that affect subhalos while orbiting the MW-like
halo.  Fig.~\ref{dist_of_angularmoment_orientation_accminus1} shows
the orientation of the angular momentum of subhalos accreted in the
last 13 snapshots, from the present-day (top left) to $z \sim 1.08$
(bottom left).  Here the angular momentum is calculated using the
position and velocity of a subhalo in the simulation box frame right
before it was accreted.  Note that only a fraction of these subhalos
will have survived until the present-time.  The small scale clustering
visible in this figure once again highlights the group infall.  Note
the presence of larger-scale patterns lasting over several snapshots
(in particular in the top row, which corresponds to the last 2.4
Gyr). This presumably implies that the infall patterns are related
with persistent larger scale structures (filaments) in the tidal
field.

\begin{figure}
   \centerline{\includegraphics[width=0.5\textwidth]{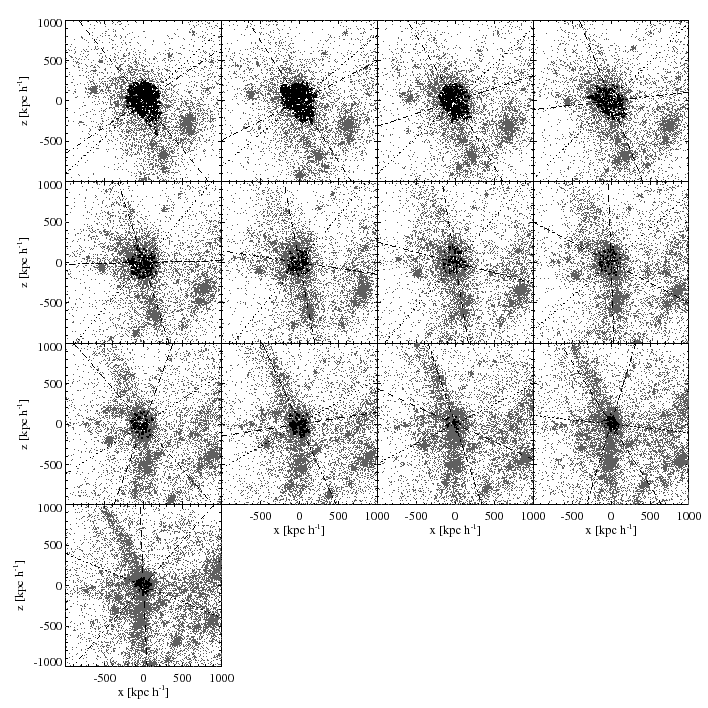}}
   \caption{Evolution of the tidal field around the Milky Way-like
halo. The grey dots represent 0.05\% of the ``field'' particles within
a box of $2h^{-1}$Mpc on a side in the simulation reference
frame. Like in
Fig.~\ref{dist_of_angularmoment_orientation_accminus1} the top left
panel corresponds to the present-time, while the bottom left to $z
\sim 1.08$. The black circles denote the location of the subhalos
accreted at the given epoch. The dashed, dash-dot and dotted lines
indicate the major, intermediate and minor axes of the MW-like halo at
each epoch.}
\label{tidal_fields}  
\end{figure}

To understand this in more detail, we proceed to trace the evolution
of the tidal field around the main halo in our simulation.  To this
end we select ``field'' particles, i.e. those that do not belong to
the FOF group of the Milky Way-like halo.  The projected spatial
distribution of these particles within a $2h^{-1}$Mpc on a side box
is shown in grey in Fig.~\ref{tidal_fields}.  Like in
Fig.~\ref{dist_of_angularmoment_orientation_accminus1}, each panel
corresponds to a different redshift, starting from $z \sim 1.08$ in
the bottom left panel to the present day in the top left.  The
distributions of surviving subhalos accreted at the corresponding
epoch are overplotted in black. 

Fig.~\ref{tidal_fields} shows that the Milky Way like halo is
embedded in a larger-scale filamentary pattern. These filaments are
comparable in extent to the halo itself (as e.g. traced by the
accreted subhalos). The lumpy nature of the filaments is also clearly
visible, showing that the infall is not a continuous flow, but is in
groups as discussed above. 

Note that the global orientation of the tidal fields near the main
halo has not changed much over the last four snapshots, in agreement
with what is observed in the top row of
Fig.~\ref{dist_of_angularmoment_orientation_accminus1}.  Furthermore,
this large scale pattern is more or less aligned with the major axis
of the main halo, shown by the dashed line in each panel \citep[as
in][]{bs05}. 

%
%
%
\subsection{On the Great Disk of Milky-Way Satellites}
\label{great_disk_sec}  

\citet{kroupa05} and \citet{metz07} have recently argued that the
highly anisotropic distribution of MW satellites could not have been
drawn from a nearly spherically distributed subhalo
population. Motivated by their claim and the results presented above,
we wish to test here under what conditions such a configuration is
likely in a $\Lambda$CDM simulation like ours.

There are many possible ways to define the degree of flattening of a
distribution. We shall here concentrate on the following two measures:
\begin{enumerate}
\item The minor-to-major axis ratio $c/a$ derived from the eigenvalues
of the diagonalised inertia tensor defined by the satellites
positions.
\item The rms of the distances to the best fit plane to the satellites
positions normalized by their median distance from the centre:
$\Delta = D_{rms}/R_{med}$ \citep{kroupa05,zentner05}.
\end{enumerate}
In what follows the positions are defined with respect to the centroid
of the satellites (or subhalos), rather than with respect to the
centre of the MW(-like) halo. For the first measure (i), we use the
inertia tensor defined as
\begin{equation}
I^\star_{ij}=\sum_{\mu} x_{i}^{\mu}x_{j}^{\mu}.
\label{inertia_tensor_eq}
\end{equation}
Note this inertia tensor differs from that previously used in
Eq.~(\ref{inertia_tensor_ell_eq}) in which the positions were
normalized by their ellipsoidal distance. Our preference for this new
definition is based on the fact that the determination of the
ellipsoidal distance is simultaneous to the determination of the
eigenvalues of the inertia tensor $I_{ij}$. This means that an
iterative algorithm is used, in which outliers are successively
discarded, until the desired level of convergence is reached (see
Sec.~\ref{group_infall_sec}). However, when a relatively limited
number of data points is available (as in the case of the MW
satellites) this is clearly not desirable.  Note as well that the
shape of $I^\star_{ij}$ is more sensitive to objects at large
distances. However, the effect this has on the measured $c/a$ can be
quantified, as we shall see below.

In our analysis, we will only consider the eleven ``traditional''
Milky Way satellites. This enables us to make a direct comparison to
the work of \citet{kroupa05}. On the other hand, it ensures we are not
affected by the strong observational bias in the sky distribution of
the new satellites discovered by SDSS, which reflects the fact that
this survey has concentrated on the north galactic cap.

The minor-to-major axis ratio for the set of eleven ``traditional'' MW
satellites is $c/a \sim 0.18 \pm 0.01$ where the uncertainty is due to
errors in the Galactocentric distance (which also includes the
uncertainty in the distance from the Sun to the Galactic centre,
$R_{\odot}=8.0\pm 0.5$ kpc).  The plane containing the major and
intermediate axes is inclined $72.8\pm 0.7^{\circ}$ with respect to
the Galactic disk, in good agreement with the value found by
\citet{metz07}.

If we use measure (ii) to quantify the degree of flattening, we find that
the best-fit plane of the MW satellites has an orientation identical
to that of the inertia tensor. The rms distance to this plane is
$D_{rms} \sim 18.5$ kpc and the plane is offset from the Galactic
centre by $7.83$ kpc.  The median distance of the satellites is $\sim
80$ kpc, which implies that $\Delta=0.23 \pm 0.01$ assuming Gaussian
errors of all related distances \citep[see also][]{metz07}.
\subsubsection{Overall distribution of subhalos}

The subhalos in our GAnew simulations show at least two differences in
their distributions in comparison to the Galactic dwarfs: 1) their
spatial distribution is much less anisotropic; 2) their density
distribution is much shallower. 

The present-day subhalo population of the Milky Way-like halo has a
minor-to-major axis ratio $c/a \sim 0.77$, obtained using
Eq.~(\ref{inertia_tensor_eq}), compared to $c/a = 0.74$ for the
MW-like halo itself. These values are fairly consistent with those
published by \citet{zentner05} and \citet{libeskind05}.  The minor to
major axis ratio for those subhalos within $\sim 300$ kpc is $c/a \sim
0.86$.  All these values are significantly larger than the $c/a \sim
0.18$ found for the Milky Way satellites.

\begin{figure}
   \centerline{\includegraphics[width=0.5\textwidth]{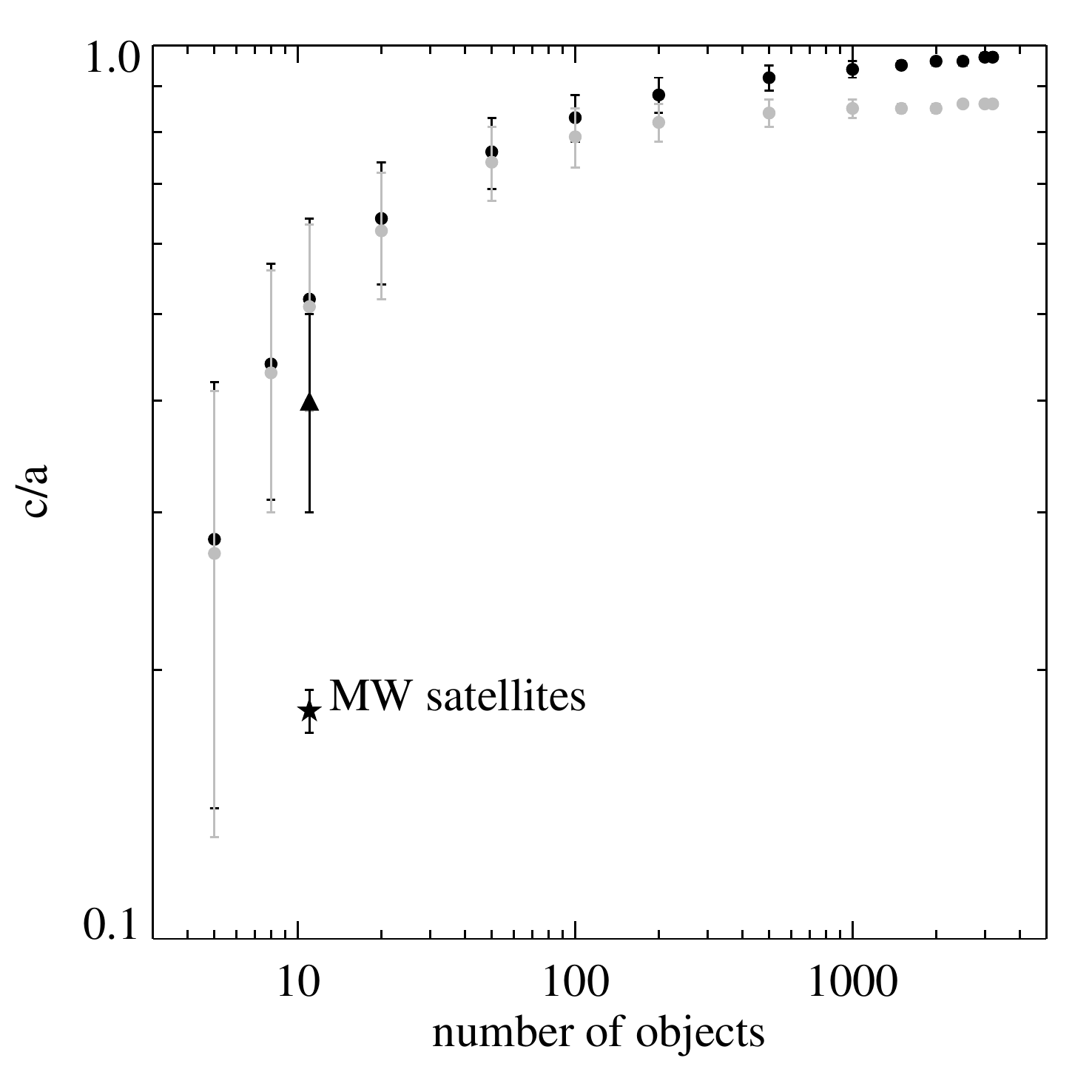}}
   \caption{Minor-to-major axis ratio of the inertia tensor defined by
random sets containing a varying number of subhalos. The grey dots are
for present-day subhalos within $300$ kpc in GA3new, while the black dots
have been drawn from an isotropic distribution. The error bars show
the $1-\sigma$ level of each distribution.  The $c/a$ obtained by
imposing that $11$ subhalos follow the MW satellites radial distribution
is denoted by the triangle, while the star symbol corresponds to the
$c/a$ for the $11$ MW satellites.}
\label{under_sampling_of_axis_ratio}
\end{figure}

We now wish to test the effect of small number statistics in the
determination of the principal axes of the inertia tensor. To this end
we randomly select 10$^5$ samples of $N$ subhalos: {\it i)} from the
simulated Milky-Way like halo population; {\it ii)} from an isotropic
distribution with the same radial profile as found in the simulation.
In Fig.~\ref{under_sampling_of_axis_ratio} we plot the mean $c/a$ as
a function of the number $N$ of subhalos selected.  The grey dots
correspond to those drawn from present-day subhalo population (case
$i$) while the black points are drawn from an isotropic sample (case
$ii$).  The error bars denote the standard deviations of the $10^{5}$
realisations. This figure shows that only when $N \sim 10^3$ a
reliable estimate of the shape of the parent distribution can be
obtained. The mean $c/a$ tends to become smaller, i.e. the
distribution appears more flattened as the number of subhalos becomes
smaller. This is also true when the parent distribution is completely
isotropic. When only $11$ subhalos are selected, the mean $c/a \sim 0.51
\pm 0.12$ ($c/a \sim 0.52 \pm 0.12$ for the isotropic case), quite
different from that of the parent sample. Nevertheless, this is still
significantly larger than observed for the MW satellites.

We noted previously that the density distribution of the satellites of
the Milky Way is strongly centrally concentrated. The median radius of
the satellite's distribution is $\sim 80$ kpc, comparable to the
half-mass radius of the Milky Way ($0.29\,r_{vir}$ for a concentration
of $c = 18$ as in \citealt{battaglia05}). On the other hand, the median
radius for the subhalos (within $300$ kpc) in our simulation is much
larger, $R_{med} \sim 0.64\,r_{vir}$.

We now quantify the effect of this highly centrally concentrated
distribution on the shape of the inertia tensor. We proceed by
randomly selecting sets of $11$ subhalos from our simulation, but now
imposing they follow the observed MW satellites spherically averaged
spatial distribution.  The mean minor-to-major axis ratio obtained in
this way is $c/a \sim 0.40 \pm 0.10$, and is denoted as a triangle in
Fig.~\ref{under_sampling_of_axis_ratio}. This exercise shows that
the disk-like configuration of the MW satellites is indeed partially
driven by their strongly centrally concentrated density distribution
around the Galaxy \citep[see also][for a similar discussion]{kang05,zentner05}. 
However, this is still only marginally consistent with
the MW satellites, i.e. the mean $c/a$ of $10^{5}$ 11-random-subhalo
realisations is $\sim 2.2 \sigma$ away from that observed.

\subsubsection{Distribution of grouped subhalos}

Inspired by the group infall of subhalos found in
Sec.~\ref{group_infall_sec}, we now explore how the presence of such
groups of subhalos can affect the chance of obtaining a flattened
configuration. Naively, we would expect that if their angular momenta
is nearly conserved, the subhalos would spread along their orbit and
give rise to a planar structure as observed.

Given the large sensitivity of $c/a$ to sample size, we prefer to use
the $\Delta$ measure in what follows. This measure intuitively appears to
be more robust since it derives from a simple plane-fit to a
distribution of points.

Of the $3,246$ subhalos within $300$ kpc from the centre of the MW-like
halo that have survived until the present-day, $898$ subhalos fell in as
part of a group. $321$ different groups have contributed to the
present-day population of subhalos, of which the earliest two were
accreted at $z=3.05$. From now on, subhalos identified to be part of a
group are referred to as ``grouped'', while those which are not are
termed ``field'' subhalos.

\begin{figure}
\hspace{-0.3cm}
\centerline{\includegraphics[width=0.52\textwidth]{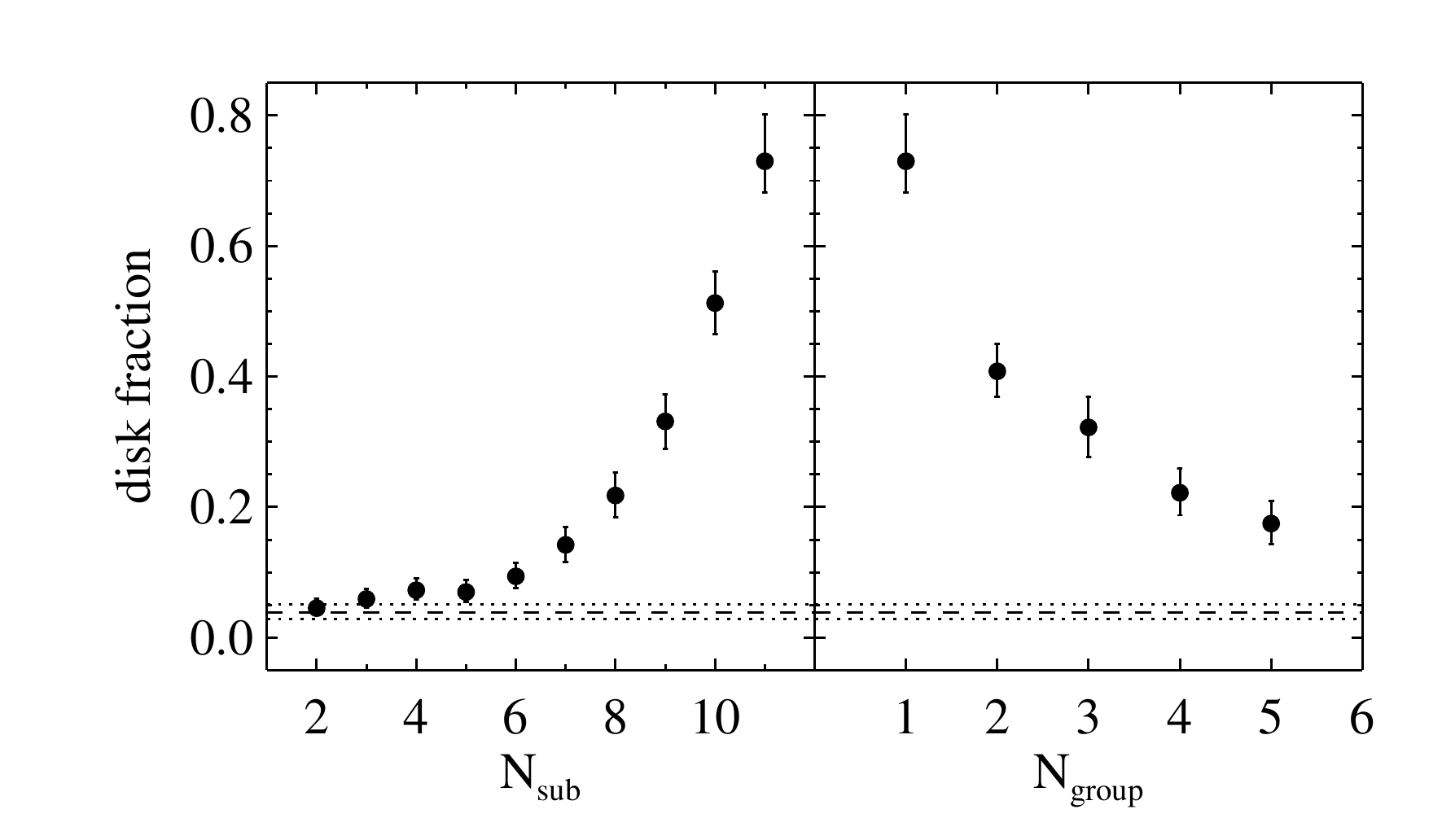}}
   \caption{{\it Left:} Fraction of disk-like structures obtained in
$10^{5}$ realisations consisting $N_{sub}$ subhalos extracted from one
group and $11-N_{sub}$ from the ``field''. {\it Right:} Fraction
obtained when $11$ subhalos are extracted from $N_{group}$ different
groups.  The fraction increases as more subhalos from one group are
selected, and as the number of groups contributing decreases. The
likelihood of obtained a highly flattened distribution is always
higher than when $11$ subhalos are randomly selected (dashed lines).}
\label{diskness_from_1grp}
\end{figure}

We first test how often a disk-like structure is obtained when
selecting a set of $11$ subhalos consisting of a certain number
$N_{sub}$ of subhalos from one group ($N_{group} = 1$) while the rest
$N_{field}$ are from the ``field''.  Note that $N_{sub} = 2, 3,
\ldots, 11$, and the condition $N_{sat} = N_{field} + N_{sub} =
11$ has to be satisfied.  We make $10^{5}$ realisations of such
subsets and compute the fraction that gives rise to a configuration as
flat as observed, i.e. $\Delta \le 0.23$.  The result is shown in the
left panel of Fig.~\ref{diskness_from_1grp} where the disk fraction
increases from 4.5\% to 73\% as the number of selected subhalos
$N_{sub}$ increases from $2$ to $11$.  In comparison, 11-randomly selected
subhalos (within $300$ kpc) gives rise to flattened configurations $\sim
2.2\%$ of the time.  This shows that if the Milky Way satellites fell
in together, it would not be very surprising that they would be in a
planar configuration at the present-day.

It is important to note that when $N_{sub} \ge 6$, we select
predominantly from just two groups accreted at relatively high
redshift ($z=1.08$ and $z=0.84$). Other large groups accreted more
recently are still strongly clustered in space, and hence are
discarded in this exercise since they cannot be considered as a valid
representation of the MW satellite population. Furthermore, although
there is a relatively high chance of obtaining a value of $\Delta$ as
low as observed, this is driven more by the large median distance of
the subhalos than by their RMS distance to the best fit plane.

A second possibility is to consider only ``grouped'' subhalos.  That
is, we select randomly $11$ subhalos from $N_{group}$ different groups
where $N_{group} = 1...5$.  The panel on the right of
Fig.~\ref{diskness_from_1grp} shows the fraction of disk-like
configurations obtained in this way as a function of the number of
groups considered.  This fraction can be as high as $\sim 40\%$ when
the subhalos come from only two groups, and of course reaches 73\%
when they come from just one group.  Note that the fraction when
selecting from $5$ different groups is still much higher than if one
selects $11$ subhalos randomly.

\begin{figure}
\hspace{-0.3cm}
   \centerline{\includegraphics[width=0.52\textwidth]{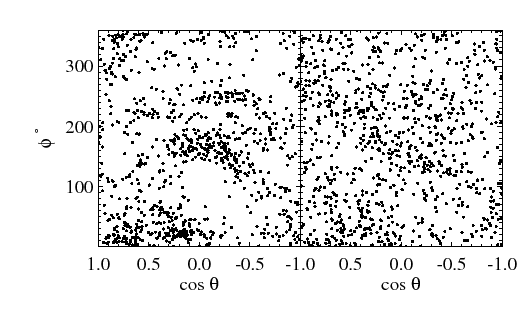}}
\caption{Present-time angular momentum orientation of subhalos located
within $300$ kpc from the MW-like host.  The left panel shows the
subhalos from ``groups'' while the right corresponds to (a randomly selected 
subset from) the ``field''.}
\label{orbitalpoles_of_grp_nongrp}
\end{figure}

The reason for the larger number of disk-like configurations when
selecting subhalos from several groups as in the right of
Fig.~\ref{diskness_from_1grp}, rather than from just one group and
the field, may be understood by examining
Fig.~\ref{orbitalpoles_of_grp_nongrp}.  This shows the present-day
angular momentum orientations for ``grouped'' (left panel) and
``field'' (right panel) subhalos. The two distributions differ clearly
in the sense that the ``grouped'' subhalos are generally more
clustered (see also Sec.~\ref{group_infall_sec}), also on larger
scales.  On the contrary, ``field'' subhalos tend to have their
angular momenta more isotropically distributed. Therefore, when
selecting $11$ subhalos purely from groups, the chance of picking up
subhalos with similar angular momentum orientations, is higher than
when selecting also from the ``field''. The more isotropically
distributed orbits of ``field'' subhalos essentially add noise to the
highly correlated orbits of subhalos originating in just one group.
Therefore the disk signal gets smeared out quite significantly when
more than half of the subhalos are in the field in one realisation, as
shown on the left of Fig.~\ref{diskness_from_1grp}.

Given the large fraction of flattened configurations found in our
simulations, we conclude that the spatial distribution of the $11$ Milky
Way satellites can be reproduced within $\Lambda$CDM. The requirement
is that these satellites fell onto the Galactic halo in groups.
%
%
%
\section{Discussion and conclusions}
\label{discussion_sec}

We have analysed a high-resolution cosmological simulation of the
formation of a Milky Way-like halo, focusing on the properties of
the satellite population at the present-day. 

We have found that dark-matter subhalos are often accreted in groups
in our simulations. Roughly $1/3$ of the surviving subhalos with mass
$\ge 2.9 \times 10^6 \msun$ at the present epoch share this
property. This is clearly a lower limit since we are not be able to
identify accompanying halos below our resolution limit (this is
particularly severe at high-redshift).

This group infall is apparent as an enhancement in the number of
subhalos whose angular momentum orientation is similar, particularly
at the time of infall. This signal is measurable also from the
present-day angular momentum of subhalos, even for those accreted 8
Gyr ago. These groups of subhalos share coherent orbits which can be
traced back well before the accretion epoch. The differential group
mass function follows a power-law distribution $dN/d \log M|_{group}
\propto M^{n}$ with $n \sim -0.5 \pm 0.2$. This is reminiscent of the
differential mass function of subhalos in both galaxy and cluster-size
halos, albeit with a shallower slope (compared to $n \sim -0.8$ as in
e.g. \citealt{delucia04,gao04b}).

We have also studied the degree of flattening of the spatial
distribution of subhalos in our simulation. The mean minor-to-major
axis ratio $c/a$ of the inertia tensor defined by the positions of $11$
randomly selected subhalos with $300$ kpc is $c/a \sim 0.51 \pm
0.12$. In comparison the $c/a$ of the $11$ ``classical'' MW satellites
is $0.18 \pm 0.01$.  Imposing the centrally concentrated MW satellite
radial distribution leads to $c/a \sim 0.4 \pm 0.1$ and therefore
somewhat alleviates the discrepancy with the observations \citep[see also][]{kang05}.

 We have explored also how this planar configuration may be obtained
as a result of the infall of satellites in groups. The observed
correlation in the angular momentum orientation of subhalos naturally
gives rise to disk-like configurations. For example, we find that if
all subhalos are accreted from just one group, it is almost impossible
to avoid a disk-like distribution ($\sim$80\% probability), while for
accretion from just two groups, the likelihood of obtaining a
distribution as planar as observed is 40\%.

These results may explain the origin of the ghostly streams proposed
by \citet{lyndenbell95}. Out of the streams originally proposed, only
two appear to have survived the rigour of time, after modern and
accurate measurements of proper motions have become available.
\citet{palma02} confirmed the LMC-SMC-UMi-Draco stream forms a
kinematic group whose angular momentum separation is $<
18.5^{\circ}$. More recently \citet{piatek05} ruled out with $95\%$
confident level UMi as a member using \textit{HST} proper motions. The
latest measurements of the Fornax proper motion by \citet{piatek07}
has apparently confirmed the Sculptor-Sextans-Fornax stream (although
previous measurements led to conflicting results, see 
\citealt{piatek02,dinescu04}).  If some of the luminous satellites are
embedded in dark (sub)halos that fell in together, such coherent
structures would be a naturally consequence of the hierarchical
build-up of galaxies.

In our simulations, such groups remain coherent in angular momentum
(i.e. they share similar orbital planes giving rise to great circle
streams) for approximately 8 Gyr. This implies that these groups (or
satellites) should have been accreted by the Milky Way at redshifts
$z\sim 1$ or below. 

One of the possible implications of the reality of the ghostly streams
is that its member galaxies formed and evolved in a similar
environment before falling into the Milky Way potential. This would
have implications on the (earliest) stellar populations of these
objects, such as for example, sharing a common metallicity floor
\citep{helmi06}. On the other hand, this implies that there should be
groups that have not been able to host any luminous satellites. This
would hint at a strong dependence on environment on the ability of a
subhalo to retain gas \citep{scannapieco01}, or be shielded from
re-ionization by nearby sources \citep{mcb04,weinmann07}.

Recent proper motion measurements of the Large and Small Magellanic
clouds by \citet{smc-mu}, as well as the simulations by \citet{bc05}
suggest that these systems may have become bound to each other only
recently. This would be fairly plausible in the context of our
results.  The Clouds may well have been part of a recently accreted
group \citep[see also][for the link to Leo I]{sales07} and it may not
even be necessary for them to ever have been a binary system. This may
also have implications on the computations of the past trajectories of
these systems, particularly if both are embedded in a larger common
dark-matter envelope \citep{besla07}.

Our analysis shows that the dynamical peculiarities of the Milky Way
satellites can be understood in the context of the concordance
cosmological model. Their properties must be a consequence of both the
environment as well as of the hierarchical nature of the build up of
galactic halos.

\section*{acknowledgements}
We thank Felix Stoehr for providing us with the GAnew simulations and
for the great support in dealing with their analysis; \mbox{Martin} Smith and Simon White for
stimulating discussions and suggestions.  We are grateful to Laura
Sales for a careful reading of the manuscript and for many useful
discussions. We acknowledge financial support from the Netherlands
Organisation for Scientific Research (NWO).

\label{lastpage}
\end{document}